\documentclass{jfm}

\usepackage{graphicx}
\usepackage{newtxtext}
\usepackage{newtxmath}
\usepackage[ruled,vlined]{algorithm2e}
\usepackage{natbib}
\usepackage{hyperref}
\hypersetup{
    colorlinks = true,
    urlcolor   = blue,
    citecolor  = black,
}

\newcommand{\RomanNumeralCaps}[1]
\linenumbers

\title{Adjoint-based particle forcing reconstruction and uncertainty quantification}

\author{Daniel Dom\'inguez-V\'azquez\aff{1},
  Qi Wang\corresp{\email{qwang4@sdsu.edu}}\aff{1}
 \and Gustaaf B. Jacobs\aff{1}}

\affiliation{\aff{1}San Diego State University, 5500 Campanile Dr, San Diego, CA 92182}
\begin{document}

\maketitle

\begin{abstract}
The forcing of particles in turbulent environments influences dynamical properties pertinent to many fundamental applications of particle-flow interactions. Current study explores the determination of forcing for one-way coupled passive particles, under the assumption that the ambient velocity fields are known. When measurements regarding particle locations are available but sparse, direct evaluation of the forcing is intractable. Nevertheless, the forcing for finite-size particles is determined using adjoint-based data assimilation. This inverse problem is formulated within the framework of optimization, where the cost function is defined as the difference between the measured and predicted particle locations. The gradient of the cost function, with respect to the forcing can be calculated from the adjoint dynamics. When measurements are subject to Gaussian noise, samples within the probability distribution of the forcing can be drawn using Hamiltonian Monte Carlo (HMC). The algorithm is tested in the Arnold-Beltrami-Childress (ABC) flow as well as the homogenious isotropic turbulence. Results demonstrate that the forcing can only be determined accurately for particle Reynolds number between 1 and 5, where the majority of Reynolds number history along the particle trajectory falls in.
\end{abstract}


%%%%%%%%%%%%%%%%%%%%%%%%%%%%%%%%%%%%%%%%%%%%%%%%%%%%%%%%%%%%%%%%%%%%%%%%%%%%%%%%
%						    	INTRODUCTION
%%%%%%%%%%%%%%%%%%%%%%%%%%%%%%%%%%%%%%%%%%%%%%%%%%%%%%%%%%%%%%%%%%%%%%%%%%%%%%%%
\section{Introduction}
The physics of dispersed particle fields in unsteady and turbulent flow fields is intricate. This has been and continues to be the inspiration of a significant amount of research on the  modeling of particle-laden flow. While several textbooks such \cite{CST98, marchisio2007multiphase, Ishii75} and review articles \cite{brandt2022particle} have summarized a large number of approaches and an even larger number of models,
a key issue in constructing robust and accurate computational models for the dynamic behavior of the dispersed particle fields remains the formulation of accurate particle forcing models.
The current research aims at establishing the methodological framework for adjoint-based inverse modelling of the particle forcing dynamics with uncertainty quantification when sparse, noisy measurements data along the particle trajectories is available.

\subsection{Particle Forcing Models}
The flow forcing of a single smooth spherical object in uniform flow is well-known and described analytically for low Reynolds and Mach number by the Maxey-Riley equation \citep{maxey1983equation}. Resolution requirements and the complexity of the interaction between wakes, boundary layers, turbulence and shocks, however, limit the accurate measurement and prediction for flow conditions outside of that range.  At higher Reynolds numbers, for example, the flow separates, and can become unstable and the forcing depends on subtle linear and non-linear instabilities that can be sensitive to small changes in the flow conditions. 
%Often high-Reynolds number effects  are lumped into empirical corrections  that while well-established, only have limited application to a single particles and they do not account for unsteady effects  (for an extensive discussion of these effects, we refer the interested reader to one of the textbooks mentioned above). These empirically corrected models are widely used and often for problems for they were not intended.
For scenarios with high-Reynolds number effects,
for example, simulations of explosively dispersed particle fields originating from a high-speed moving source, basic empirical forcing models have been used \citep{boiko1997shock, jacobs2009high, jacobs2012high}.
However, it is clear that the accuracy of such models is affected by unsteady forcing \citep{sen2018role} and unsteady carrier-phase effects \citep{sen2019machine}. In fact, high-speed particle flow has spurred model development over the last decade \citep{Bala, capecelatro2014numerical, jacobs2012high} and led to studies that determine either theoretically, experimentally or through machine learning the particle forcing of single particles affected by moving shocks 
\citep{KKV06,ling2011importance,parmar2009modeling} and arrays of particle affected by shock diffraction  \citep{sen2017evaluation, ling2012interaction,mehta2016shock, capecelatro2014numerical, tryggvason2013direct, regele2012numerical,regele2014unsteady}. 

In addition to the impact of flow conditions, a range of geometric parameters can affect the particle. For example micro- and meso-scopic, but crucial geometric imperfections and deformation of (condensed) particles’ surfaces impacts on momentum and energy exchange
with a carrier flow. A harmonic perturbation of cylindrical shape for example can change the dominant vorticity generation  by shear to a baroclinic mechanism \citep{blanco2022wall} and yield different forcing. Just these few examples show how many parameters and flow conditions can affect forcing models, and it should not be surprising that an accurate and comprehensive forcing model has eluded the community and realistically is not be feasible
for all conditions.

As an alternative to understanding the effect of more flow conditions (free-stream turbulence, free-stream velocity profile, stratification etc.) or particle geometry and the formulation of even more additional and new forcing corrections, we might consider the forcing problem from an inverse perspective and resolve the limitations of modeling and experiment through inference and optimization from observational and simulated data using analytical relations and inputs and outputs of simulation and experiment.
Could we, for example, infer the forcing from limited trajectory data? Or might it be possible to assume a random forcing models with a confidence interval and improve this forcing with a limited amount of high-resolution data?  
To this end an intelligent framework is required that systematically and robustly improves forcing models and associated macro-response protocols with a robust uncertainty quantification using limited, experimental data.
% To this end, we propose to develop a transformative framework based on the digital twin paradigm for the
% modeling of the time-dependent interaction between the motion of particles and hypersonic flow dynamics
% over a wedge. The framework combinesstate-of-the-art physics-based stochastic particle and particle-group
% models and sparse regression techniques with experimental measurements in a supersonic blow down tunnel through adjoint analysis.

% \gbj{Qi: here a review of inverse methods}
\subsection{Inverse methods}
The inference of particle forcing from sparse measurements can be accomplished from an inverse-problem perspective. 
The dependence of the measurements, either individual particle trajectory data or shape of the cloud distribution, onto the forcing law is highly nonlinear, obtained from solving the physical governing equations.
In general, for such nonlinear physics-constrained inverse problems, two different classes of methods are considered in past researches. 
Data-based methods aim at learning massive amount of data as a distribution, often in reduced-order latent space. Methods such as Generative adversarial networks (GANs) \citep{hassanaly2022adversarial,silva2021data,silva2021gan,legler2022combining}, Auto-encoders \citep{wang2022deep,peyron2021latent,amendola2020data} or Long-Short-Term Memory (LSTM) \citep{wan2018machine} can successfully discover the intrinsic coordinates of the data and take into account the effect of observing by evaluating the probability distribution conditioned on those observations.
However, giving the efficiency and practicality of these methods, a thorough physical analysis of the results is often difficult because the foundation is not the governing equation.
Model-based methods, on the other hand, initiate from the governing equation and often need to solve the adjoint equations to obtain the gradient of the cost function \citep{wang2019spatial}.
To maintain consistency with the forward solver and ensure numerical stability during optimization, discrete adjoint operator is often derived and implemented \citep{wang2019discrete}. 
Furthermore, the adjoint fields by solving the adjoint equations gives clear interpretation of the physical meaning \citep{wang_wang_zaki_2022} as the domain of dependence.

One significant benefit of data-based methods is that uncertainty quantification is straightforward under the probabilistic framework.
However, for model-based methods, uncertainty quantification is often much more difficult in terms of the adjoint dynamics. 
For linear dynamics, Bayesian inference with uncertainty quantification can be easily performed using Gaussian Process Regression (GPR) through the aid of adjoint \citep{gahungu2022adjoint}.
For nonlinear dynamics, distribution of the control vector deviates from Gaussian and samples of the probability distribution can be draw from Hamiltonian Monte Carlo (HMC) using the gradient Langevin dyanmics \citep{yang2021b, herrmann2019learned}.
A remaining hassle would be the computational cost to solve both the forward and adjoint governing equations repetitively.

To account for parametric, empirical and structural uncertainty fluctuations in forcing models, a probabilistic (stochastic), point-particle and multi-scale perspective as was taken in \cite{jacobs2019uncertainty} and \cite{dominguez2021lagrangian}. Starting from known empirical models and/or surrogate forcing models from high-resolution simulations \citep{sen2018evaluation}, that
account for Reynolds, Mach number and number density, and other parameters, we can describe the forcing and its dependencies within confidence intervals according to a probability density function. Stochastic dynamic models are formulated that propagate this distribution into the stochastic solutions of trajectories \citep{dominguez2021lagrangian}. 

In this paper, we reconstruct forcing models using a combination of governing dynamic equations that govern the location and velocity of a particle modeled
as a singular point with an unknown forcing and adjoint formulations.
We develop a theoretical framework that determines forcing for one-way coupled passive particles, under the assumption that the ambient velocity fields are known.
%When measurements regarding particle locations are available but sparse, directly evaluation of the forcing is intractable. 
%Using the probablistic models, we compared trajectories with quantified uncertainty and compared to data points from experiment and,
%using those data points, the governing equations can be systematically improved through techniques from the field of optimization. 
%We will assume point-particle, i.e., the particle location and its forcing is assumed at a single, singular point. Of course the data can that is used to converge on this
%forcing may stem from  a finite-size particle geometry is determined using adjoint-based data assimilation. 
%By defining a  cost function  as the difference between the measured and predicted particle locations, we can propose to compute gradient of the cost function, with respect to the forcing can be calculated from the adjoint dynamics. 
%When measurements are noisy, uncertainty quantification regarding the forcing can be evaluated using Hamiltonian Monte Carlo.

The remainder of the paper is organized as follows: firstly we focus on point-particle and particle forcing models and derive the optimization framework based utilizing the adjoint dynamics. The forcing reconstruction is then practiced in two test cases, the Arnold–Beltrami–Childress (ABC) flow and a decaying isotropic turbulence flow.

\section{Problem Formulation}
\label{Sec:ProblemSetup}

\label{Sec:numericalmethods}
Consider the motion of a single particle with trajectory $\boldsymbol{x}_p(t)$ and velocity $\boldsymbol{u}_p(t)$.
The non-dimensional governing equations for the particle read
\begin{equation}
\begin{aligned}
    \frac{d \boldsymbol{x}_p}{d t} = \boldsymbol{u}_p, \quad  \frac{d \boldsymbol{u}_p}{d t} = \frac{f}{St}\left(\boldsymbol{u} - \boldsymbol{u}_p\right),
\end{aligned}
\label{eqn:GE}
\end{equation}
where $\boldsymbol{u}$ is the ambient velocity at the particle location $\boldsymbol{x}_p$.
The Stokes number is defined as the ratio between the characteristic time of the flow and the particle phase, and can be expressed in terms of the reference Reynolds of the flow as $St = Re_\infty\frac{\rho_p d_p^2}{18}$ where $d_p$ is the non-dimensional particle diameter and $\rho_p$ is the density ratio of the two phases.
The function 
$f$ is the forcing coefficient as a fix for Stokes flow.
For Stokes flow, $f$ becomes the unity function. 
In more general cases, the forcing term $f$ is a function of the slip velocity, $f = f(|\boldsymbol{u}-\boldsymbol{u}_p|) = f(a)$.

The governing equation is integrated using simple forward Euler scheme, with two ambient velocity fields considered in the current study: the Arnold–Beltrami–Childress (ABC) flow and the decaying homogeneous isotropic turbulence obtained from Johns Hopkins Turbulence Database (JHTDB) \citep{li2008public}.
computed with an in-house discontinuous Galerkin (DG) DNS solver described in~\cite{klose2020assessing} and references therein, with initial conditions adopted from~\cite{jacobs2005validation}.

In principle, $f$ lives in the infinite dimensional space. In the current study, we adopted one of the most common forcing models~\citep{boiko1997shock,jacobs2012high} as the ground truth,
\begin{equation}
    f(a) = \left( 1 + 0.38\frac{Re_p}{24} + \frac 1 6 \sqrt{Re_p} \right),
    \label{eqn:trueforcing}
\end{equation}
with the particle Reynolds number defined as  $Re_p = Re_\infty a d_p$.

In computations, proper low-rank approximation of the forcing function has to be applied, and the forcing is discretized as a linear superposition of orthogonal basis functions,
\begin{equation}
\label{eqn:Fourier}
    f(a) = \sum_i \alpha_i \psi_i(a),\quad \psi_i(a) = \cos \left(i\frac{2\pi d_p Re_{\infty}}{2 Re_{p,max}} a\right),
\end{equation}
where $Re_{p,max}$ is set to be slightly more than the maximum particle Reynolds number encountered in the simulation.
We denote the parameters $\boldsymbol{\alpha}$ for dimensional reduction.
The determination of $f$ is then reduced to the determination of the parameters $\boldsymbol{\alpha}$.

% and we used the following Fourier basis functions for the discretization of the forcing,
% \begin{equation}
% \label{eqn:Fourier}
%     \psi_i(a) = \cos ((i-1)\frac{2\pi}{2A} a),
% \end{equation}
% where $A$ is set to be the maximum velocity difference encountered in the simulation.

\subsection{Adjoint optimization algorithm}
When sparse measurements for the particle trajectory are available, information regarding the forcing is indirect. 
We consider a special case, where the measurement is only available at the final time $t_m$. 
The cost function in this case can be defined as the difference between the computed and measured particle locations at $t_m$, namely,
\begin{equation}
    J = \frac 12 ||\boldsymbol{x}_m - \boldsymbol{x}_p(t_m)||^2.
    \label{eqn:cost}
\end{equation}

For general cases where measurements are available for several different $t_m$, a simple summation of the individual cost functions will be adopted and a similar procedure is still valid.

To satisfy the governing equations \ref{eqn:GE} in addition to minimize the cost function \ref{eqn:cost}, we augment the cost function with the Lagrange multiplier $\boldsymbol{x}^\dagger_p, \ \boldsymbol{u}^\dagger_p$, which results in the function $\mathcal{H}$,
\begin{equation}
\begin{aligned}
\mathcal{H} &= J + \int_t \left[\boldsymbol{x}^{\dagger}_p\cdot \left(\frac{d \boldsymbol{x}_p}{dt} - \boldsymbol{u}_p\right) dt+ \boldsymbol{u}^{\dagger}_p\cdot\left( \frac{d \boldsymbol{u}_p}{d t} - \frac{f}{St}\left(\boldsymbol{u} - \boldsymbol{u}_p\right)\right)\right]dt.
\end{aligned}
\end{equation}

If the forcing parameters $\boldsymbol{\alpha}$ are perturbed by an amount $\delta \boldsymbol{\alpha}$, the change of forcing would result in a deviation of $\boldsymbol{x}_p$ and $\boldsymbol{u}_p$ from its original trajectory, perturbed by the amount $\delta \boldsymbol{x}_p$ and $\delta \boldsymbol{u}_p$, respectively.
In particular, the perturbation of the ambient velocity $\boldsymbol{u}\big\rvert_{\boldsymbol{x}_p}$ can be obtained from 
%Taylor's expansion as follows,
%\begin{equation}
%    \boldsymbol{u}\big\rvert_{\boldsymbol{x}_p + \Delta \boldsymbol{x}_p} =  \boldsymbol{u}\big\rvert_{\boldsymbol{x}_p} + \nabla \boldsymbol{u}\big\rvert_{\boldsymbol{x}_p} \cdot \delta \boldsymbol{x}_p + \mathcal{O}(||\boldsymbol{x}_p||^2).
%\end{equation}
%Therefore, 
\begin{equation}
    \delta \boldsymbol{u}\big\rvert_{\boldsymbol{x}_p} = \nabla \boldsymbol{u}\big\rvert_{\boldsymbol{x}_p} \cdot \delta \boldsymbol{x}_p.
\end{equation}
In addition, the change of forcing $f(a)$ can be evaluated as,
\begin{equation}
\begin{aligned}
    \delta [f(a)] &= \delta [\sum_i \alpha_i \psi_i(|\boldsymbol{u}-\boldsymbol{u}_p|)] \\&= \sum_i \delta\alpha_i \psi_i(|\boldsymbol{u}-\boldsymbol{u}_p|) + \frac{f^{\prime}(a)}{a} \left(\nabla \boldsymbol{u}\big\rvert_{\boldsymbol{x}_p} \cdot \delta \boldsymbol{x}_p - \delta \boldsymbol{u}_p\right)\cdot \left(\boldsymbol{u}-\boldsymbol{u}_p\right) 
\end{aligned}
\end{equation}
The perturbation of the cost function can be evaluated as,
\begin{equation}
\begin{aligned}
\delta \mathcal{H} =& \frac{\partial J}{\partial \boldsymbol{x}_p} \delta \boldsymbol{x}_p + \frac{\partial J}{\partial \boldsymbol{u}_p} \delta \boldsymbol{u}_p + \int_t\boldsymbol{x}^{\dagger}_p\cdot \left(\frac{d \delta \boldsymbol{x}_p}{dt} - \delta \boldsymbol{u}_p \right)dt 
\\ &  + \int_t \boldsymbol{u}^{\dagger}_p\cdot\Bigg\{ \frac{d \delta \boldsymbol{u}_p}{d t} - \frac{1}{St}f(a)  \left(\nabla\boldsymbol{u}\cdot \delta  \boldsymbol{x}_p- \delta \boldsymbol{u}_p\right) - \frac{1}{St}\delta [f(a)]\left(\boldsymbol{u} - \boldsymbol{u}_p\right)\Bigg\}dt  .
\end{aligned}
\end{equation}
% \begin{equation}
% \begin{aligned}
% \delta H =& \frac{\partial J}{\partial \boldsymbol{x}_p} \delta \boldsymbol{x}_p + \frac{\partial J}{\partial \boldsymbol{u}_p} \delta \boldsymbol{u}_p + \int_t\boldsymbol{x}^{\dagger}_p\cdot \left(\frac{d \delta \boldsymbol{x}_p}{dt} - \delta \boldsymbol{u}_p \right)dt 
% \\ &  + \int_t \boldsymbol{u}^{\dagger}_p\cdot\Bigg\{ \frac{d \delta \boldsymbol{u}_p}{d t} - \frac{1}{St}f(a)  \left(\nabla\boldsymbol{u}\cdot \delta  \boldsymbol{x}_p- \delta \boldsymbol{u}_p\right)\\
% & - \frac{1}{St}\underbrace{\left[\sum_i \delta\alpha_i \psi_i(|\boldsymbol{u}-\boldsymbol{u}_p|) + \frac{f^{\prime}(a)}{a} \left(\nabla \boldsymbol{u}\big\rvert_{\boldsymbol{x}_p} \cdot \delta \boldsymbol{x}_p - \delta \boldsymbol{u}_p\right)\cdot \left(\boldsymbol{u}-\boldsymbol{u}_p\right)\right]}_{\delta [f(a)]}\left(\boldsymbol{u} - \boldsymbol{u}_p\right)\Bigg\}dt 
% \end{aligned}
% \end{equation}
Through integration by parts, we can obtain another form,
\begin{equation}
\begin{aligned}
\delta \mathcal{H} =& \int_t\delta \boldsymbol{x}_p \cdot \underbrace{\left[-\frac{d \boldsymbol{x}_p^\dagger}{dt} - \frac{1}{St}\left(\nabla \boldsymbol{u}\right)^T \cdot \left[ f(a) \boldsymbol{u}^\dagger_p + \frac{f^{\prime}(a)}{a} \left[\left(\boldsymbol{u}-\boldsymbol{u}_p\right)\cdot \boldsymbol{u}^\dagger_p\right] \left(\boldsymbol{u}-\boldsymbol{u}_p\right)\right]+\frac{\partial J}{\partial \boldsymbol{x}_p}\right]}_{=0, \text{adjoint equations}}dt 
 \\ &+ \int_t \delta \boldsymbol{u}_p \cdot\underbrace{\left[ -\frac{d \boldsymbol{u}_p^\dagger}{d t} -\boldsymbol{x}^\dagger_p + \frac{1}{St} \left[f(a) \boldsymbol{u}_p^\dagger+\frac{f^{\prime}(a)}{a}\left[\left(\boldsymbol{u}-\boldsymbol{u}_p\right)\cdot \boldsymbol{u}^\dagger_p\right] \left(\boldsymbol{u}-\boldsymbol{u}_p\right)\right] + \frac{\partial J}{\partial \boldsymbol{u}_p}\right]}_{=0, \text{adjoint equation}}dt
 \\&- \frac{1}{St}\int_t \sum_i \delta\alpha_i \psi_i(|\boldsymbol{u}-\boldsymbol{u}_p|)\;\boldsymbol{u}_p^\dagger \cdot \left(\boldsymbol{u} - \boldsymbol{u}_p\right) dt.
\end{aligned}\label{eqn:Adjoint}
\end{equation}
% Therefore, we define the adjoint equations as,
% \begin{equation}
% \label{eqn:Adjoint}
% \begin{aligned}
%     \frac{d \boldsymbol{x}_p^\dagger}{d\tau}& =  \frac{1}{St}\left(\nabla \boldsymbol{u}\right)^T \cdot \left[ f(a) \boldsymbol{u}^\dagger_p + \frac{f^{\prime}(a)}{a} \left[\left(\boldsymbol{u}-\boldsymbol{u}_p\right)\cdot \boldsymbol{u}^\dagger_p\right] \left(\boldsymbol{u}-\boldsymbol{u}_p\right)\right]-\frac{\partial J}{\partial \boldsymbol{x}_p},\\% + \frac{2c_r}{3Pr}\frac{1}{St} f_2 T^\dagger_p\nabla T \\
%     \frac{d \boldsymbol{u}_p^\dagger}{d\tau} &= \boldsymbol{x}_p^\dagger-\frac{1}{St} \left[ f(a) \boldsymbol{u}^\dagger_p + \frac{f^{\prime}(a)}{a} \left[\left(\boldsymbol{u}-\boldsymbol{u}_p\right)\cdot \boldsymbol{u}^\dagger_p\right]\left(\boldsymbol{u}-\boldsymbol{u}_p\right)\right] - \frac{\partial J}{\partial \boldsymbol{u}_p},.
%     %\\
%     %\frac{dT^{\dagger}_p }{d\tau} &= %-\frac{2c_r}{3Pr}\frac{1}{St} f_2 T_p^\dagger.
% \end{aligned}
% \end{equation}
In case of the single measurement at $t_m$,
the derivatives of the cost function are,
\begin{equation}
\label{eqn:dJdxp}
    \frac{\partial J}{\partial \boldsymbol{x}_p} = (\boldsymbol{x}_p(t_m) - \boldsymbol{x}_m)\delta (t-t_m), \quad \frac{\partial J}{\partial \boldsymbol{x}_p} = \boldsymbol{0}. 
\end{equation}
in backward time axis $\tau = t_m-t$. 
The initial conditions of the adjoint simulations are defined as,
\begin{equation}
    \boldsymbol{x}_p^\dagger =\boldsymbol{u}_p^\dagger =  \boldsymbol{0},%\quad  T_p^\dagger = T_m - T_p
     \quad \tau = 0.
\end{equation}
Although the source term of the adjoint equation, $\partial J / \partial \boldsymbol{x}_p$ can be taken into account in the initial condition instead, we here retain the most general form so that the adjoint equations apply to a general cost function.
Also notice that the adjoint equation involves a singularity around $a = |\boldsymbol{u} - \boldsymbol{u}_p| = 0$, corresponding to the vanishing forcing when $\boldsymbol{u} - \boldsymbol{u}_p = \boldsymbol{0}$ in the forward equation.

By solving the adjoint equations backwards in time, we obtain the final gradients as,
\begin{equation}
    \frac{\partial H}{\partial \alpha_i} = - \frac{1}{St}\int_t  \psi_i(|\boldsymbol{u}-\boldsymbol{u}_p|) \boldsymbol{u}_p^\dagger \cdot \left(\boldsymbol{u} - \boldsymbol{u}_p\right) dt .
    \label{eqn:AdjointGradient}
\end{equation}
In the current study, the discretization of the adjoint equations are derived from summation by parts using the discretized forward equations. 
This approach, instead of discretizing the adjoint equations derived from integration by parts, is called the discrete adjoint \citep{mengze2019discrete}. 
The discrete adjoint calculates the gradient directions $\partial H /\partial \boldsymbol{\alpha}$ accurately to machine zero and guarantees the convergence of the optimization algorithm.

\subsection{Hamiltonian Monte Carlo}
\label{sec:HMC}
Hamiltonian Monte Carlo, which is known as a golden approach for sampling from posterior distributions, is an efficient Markov Chain Monte Carlo (MCMC) method based on the Hamiltonian dynamics.
Suppose that the posterior distribution of the unknown flow parameters, $\boldsymbol{\alpha}$, follows the distribution,
\begin{equation}
    p(\boldsymbol{\alpha}) \sim \exp\left(-\frac{J(\boldsymbol{\alpha})}{\sigma^2}\right).
\end{equation}
Given the form of the cost function $J$ being quadratic, we are sampling from the posterior distribution of the flow state given measurements being Gaussian variables.
To sample from this given posterior, HMC constructs a Hamiltonian system with a fictitious momentum variable $\boldsymbol{r}$,
\begin{equation}
    H(\boldsymbol{\alpha},\boldsymbol{r}) = \frac{J}{\sigma^2} + \frac 12 \boldsymbol{r}^T\mathbf{M}^{-1}\boldsymbol{r},
    \label{eqn:HMC_eqn}
\end{equation}
and we will sample from the joint distribution
\begin{equation}
    \pi(\boldsymbol{\alpha},\boldsymbol{r}) \sim \exp\left(-\frac{J(\boldsymbol{\alpha})}{\sigma^2}- \frac 12 \boldsymbol{r}^T\mathbf{M}^{-1}\boldsymbol{r}\right),
\end{equation}
for which we ignore the part for the momentum and obtain the marginal distribution for $\boldsymbol{\alpha}$.
The sampling is done by selecting $\boldsymbol{r}$ from its Gaussian distribution and integrate equation \eqref{eqn:HMC_eqn} using symplectic algorithms such as leapfrog. 
In addition, Metropolis-Hastings step is incorporated to enhance the efficiency of the sampling \citep{chib1995understanding}.
Algorithmically, it enables the original adjoint-based gradient descend algorithm \citep{wang_hasegawa_zaki_2019,mengze2019discrete} to extend to a probabilistic framework by adding randomness. 

The overall procedure of HMC is shown in algorithm \ref{alg:HMC}. 
The gradient of the cost function, $\nabla_{\boldsymbol{\alpha}}$ is evaluated from adjoint simulations and using equation \eqref{eqn:AdjointGradient}.
The computational cost for solving the adjoint equation is comparable to solving the forward problem and does not depend on the number of unknown parameters, e.g. the size of $\boldsymbol{\alpha}$.
In the current study, the parameters for HMC are set considering the efficiency for both the number of equation-solving and convergence behavior of the Monte Carlo sampling.

\begin{algorithm}
\DontPrintSemicolon % Some LaTeX compilers require you to use \dontprintsemicolon instead
\KwIn{A starting parameter $\alpha_0$ and step size $\delta t$}
\KwOut{Samples of the parameter $\alpha_i$ drawn from its posterior distribution.}
\For{$i \gets 1$ \textbf{to} $N$} {
  Draw $\boldsymbol{r}_{i-1}$ from $\mathcal{N}(0,\mathbf{M})$\;
  $(\tilde{\boldsymbol{\alpha}}_0, \tilde{\boldsymbol{r}}_0) \gets (\boldsymbol{\alpha}_{i-1}, \boldsymbol{r}_{i-1})$\;
  \For{$j \gets 0$ \textbf{to} $L-1$} {
    $\tilde{\boldsymbol{r}}_j \gets \tilde{\boldsymbol{r}}_j - \frac 12 \delta t \nabla_{\boldsymbol{\alpha}}\mathcal{J}(\tilde{\boldsymbol{\alpha}}_j)$\;
    $\tilde{\boldsymbol{\alpha}}_{j+1} \gets \tilde{\boldsymbol{\alpha}}_j + \delta t \mathbf{M}^{-1}\tilde{\boldsymbol{r}}_j$\;
    $\tilde{\boldsymbol{r}}_j \gets \tilde{\boldsymbol{r}}_{j+1} - \frac 12 \delta t \nabla_{\boldsymbol{\alpha}}\mathcal{J}(\tilde{\boldsymbol{\alpha}}_{j+1})$\;
  }
  Metropolis-Hasting algorithm:\;
  $p \gets \min\{1,\exp(-H(\tilde{\boldsymbol{\alpha}}_L,\tilde{\boldsymbol{r}}_L) + H(\boldsymbol{\alpha}_{i-1},\boldsymbol{r}_{i-1}))\}$\;
  Draw $\beta$ from uniform distribution in $[0,1]$\;
  \If{$\beta \leq p$}{
  $\boldsymbol{\alpha}_i \gets \tilde{\boldsymbol{\alpha}}_L$
  }
  \Else
  {
  $\boldsymbol{\alpha}_i \gets \boldsymbol{\alpha}_{i-1}$
  }
}
Use $\boldsymbol{\alpha}_i$ as samples of $\boldsymbol{\alpha}$\;
\caption{Hamiltonian Monte Carlo}
\label{alg:HMC}
\end{algorithm}

% \subsection{Behavior in the linear regime}
% When the noise is infinitesimal and the forcing is perturbed near the true solution, the perturbation of the forcing and of the measurement are related linearly, namely $\delta \boldsymbol{x}_p(t_m) = \mathcal{A} \delta \boldsymbol{\alpha}$.
% If the measured particle locations follows a Gaussian distribution with covariance matrix $\sigma^2 \mathbf{I}_3$. If the $\delta \boldsymbol{\alpha}$ has covariance matrix $\mathcal{C}_{\boldsymbol{\alpha}}$, the linear transformation between covariance matrices yields,
% \begin{equation}
%     \mathcal{A}\mathbf{C}_{\boldsymbol{\alpha}}\mathcal{A}^T = \sigma^2 \mathbf{I}_3.
% \end{equation}
% % the gradient of the cost function vanishes, its landscape can be fully described by the second order derivative, or the Hessian matrix.
% % The fact that the cost function is a quadratic form can be exploited to facilitate evaluation of the Hessian.
% % In fact, the Hessian could be obtained form the adjoint simulation, given that
% % \begin{equation}
% %     \frac{\partial^2 J}{\partial \boldsymbol{\alpha}\partial \boldsymbol{\alpha}} = \frac{\partial (\boldsymbol{x}_p(t_m))}{\partial \boldsymbol{\alpha}}\frac{\partial (\boldsymbol{x}_p(t_m))}{\partial \boldsymbol{\alpha}}
% % \end{equation}

\section{Applications}
\subsection{One-parameter estimation in the ABC flow}
% \begin{figure}
%     \centering
%     \includegraphics[width=0.7\textwidth]{ABC_differentSt.jpg}
%     \caption{Particle trajectories with different Stokes number $St = \{1,2,5,10\}$, marked by increasing line thickness.
%     The particles are released from the same initial location $\boldsymbol{x}_p(t=0) = [\pi, \pi, \pi]$. 
%     The background vector fields show the in-plane velocity fields while the colored contours are the out-of-plane vorticity component.}
%     \label{fig:ABCflow}
% \end{figure}
We first consider the Arnold–Beltrami–Childress (ABC) flow \citep{jeffreys1928some}, a three-dimensional steady analytical solution of the Euler equation, with the constants of the carrier flow field chosen as $A = \sqrt{3}$, $B = \sqrt{2}$ and $C=1$.
% which is called the ABC flow,
% \begin{equation}
% \begin{aligned}
%     u = A\sin z + C \cos y,\\
%     v = B\sin x + A \cos z,\\
%     w = C\sin y + B \cos x.
% \end{aligned}
% \end{equation}
% The ABC flow is an analytical solution of the Euler equation,
% \begin{equation}
%     \frac{\partial \boldsymbol{u}}{\partial t} + \boldsymbol{u} \cdot \nabla \boldsymbol{u} = -\frac{1}{\rho}\nabla p,
%     \quad \quad\nabla \cdot \boldsymbol{u} = 0.
% \end{equation}
\begin{figure}
    \centering
    \includegraphics[width=0.86\textwidth]{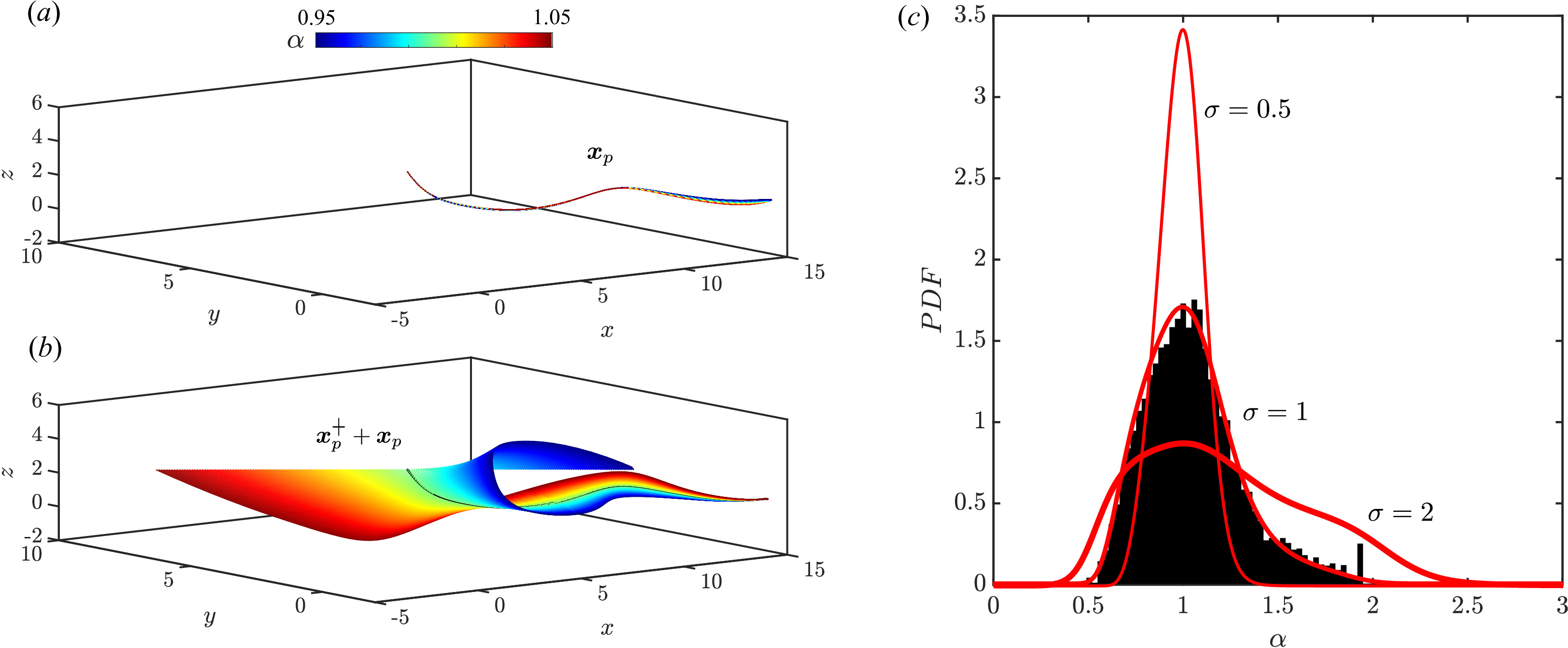}
    \caption{(a): The particle trajectories  from $t=0$ to $t = t_m = 8$ for different values of $\alpha$, while the true value is $\alpha_t = 1$.
    Increasing values of $\alpha$ from $0.95$ to $1.05$ are marked by different colors.
    (b): The adjoint particle trajectories $\boldsymbol{x}_p^\dagger + \boldsymbol{x}_p$ given the true measurement $\boldsymbol{x}_p(t = t_m)$.
    Increasing Stokes number are marked with the same color map as (a).
    (c): When the measurement data $\boldsymbol{x}_p(t = t_m)$ has Gaussian error with standard deviation $\sigma$, the posterior probability distribution of the value $\alpha$ is estimated both theoretically and by using HMC. Red lines marks the theoretical results with different level of Gaussian noise $\sigma = \{0.5, 1, 2\}$ while the histogram shows the result from HMC with $\sigma = 1$.}
    \label{fig:single_mode_UQ}
\end{figure}
The flow field is shown in figure \ref{fig:single_mode_UQ}, with particle originated at rest from $\boldsymbol{x}_{p,0} = (\pi,\pi,\pi)^\top$ with different Stokes numbers $St = \{1,2,5, 10\}$.
As a demonstration of the data assimilation and adjoint-based uncertainty quantification, we focus on a simple one-parameter reconstruction, where the forcing function is assumed to be,
\begin{equation}
    f(a) =  \alpha \psi(a), \quad\quad \psi(a) = \left( 1 + 0.38\frac{Re_p(a)}{24} + \frac 1 6 \sqrt{Re_p(a)} \right),
\end{equation}
where $\alpha$ is the only unknown parameter. Here we fix the Stokes number $St = 1$ while bringing the readers to the equivalent effect between changing $\alpha$ and changing $St$.

For $\alpha$ fluctuating in a range $0.95 \le \alpha \le 1.05$, the trajectory of the particle, $\boldsymbol{x}_p$ various slightly, as shown in figure \ref{fig:single_mode_UQ}$(a)$ in the upper panel. 
As the true value of $\alpha$ is $\alpha_T = 1$, the other perturbed values of $\alpha$ lead to mismatched observations at the time $t = t_m = 8$.
As illustrated in equations \eqref{eqn:Adjoint} and \eqref{eqn:dJdxp}, the initial position of the adjoint particles are given by the mismatch of the observation, $\boldsymbol{x}^\dagger_{p}(t_m) =  \boldsymbol{x}_m - \boldsymbol{x}_{p}(t_m) $, therefore, the trajectory $\boldsymbol{x}^\dagger_p + \boldsymbol{x}_p$ originates all from the same location $\boldsymbol{x}_m$ at the measurement time $t_m$, and travels backwards in time, forming a collection of adjoint particle tracers shown in figure \ref{fig:single_mode_UQ}$(a)$ on the bottom panel. We here stress the important properties between the forward and adjoint trajectories: nearby particles separates according to the same Lypunov exponent: a conclusion directly derived form the transposed dynamics of the adjoint \citep{wang_wang_zaki_2022}. 

The gradient of the cost function with respect to the parameter $\alpha$ is then given by the forward and adjoint particle velocities, as shown in equation \eqref{eqn:AdjointGradient}. 
For different tolerance of observation error $\sigma$, the posterior distribution of the parameter $\alpha$ can be estimated through the HMC algorithm described in section \ref{sec:HMC}. 
For this simple example with only one parameter, the PDF is also evaluated either through a direct mapping between $\alpha$ and $\boldsymbol{x}_p(t_m)$ as a validation for the HMC algorithm. Both results are shown in figure \ref{fig:single_mode_UQ}$(b)$. Results with HMC with $\sigma = 1$ shows good agreement with direct mapping. 

Results for larger value of $\sigma$ show increasing level of non-linearity: since the observation error follows a Gaussian distribution, the posterior distribution for the parameter would be perfectly Gaussian if the dynamic system is linear, which is the case for small $\sigma$; the distribution would deviate from Gaussian for larger $\sigma$.
%The results shed light on the validity of using a linearized framework such as Gaussian Process Regression (GPR) for problems with small perturbations.
%However, for finite level of observation noise, the posterior distribution has to be estimated through more complex algorithms such as HMC.

\subsection{Forcing reconstruction in the ABC flow}
As the next step, we consider the more general case where the forcing function is parametrized by a few Fourier modes, as illustrated in equation \eqref{eqn:Fourier}. 
In the application of ABC flow, where $Re_p$ is below 10, a total number of seven Fourier modes is enough to efficiently represent the shape in \eqref{eqn:trueforcing}.
The particle is initialized at location $x_p=y_p=z_p=2\pi$ at rest. The Reference Reynolds number is $Re_\infty=250$, the Stokes number is unity $St=1$, the particle density ratio is $\rho_p=500$ and the non-dimensional particle diameter $d_p=0.012$.
We assume that the initial location of the particle at rest is known, and the observations of the three coordinates of the particle location is available at time $t=8$.
We choose the observation of particle position at the non-dimensional time of eight Stokes time units so that the particle trajectory is dominated by an inertial response to the changes in the flow field.

Using the gradient of the cost function, following equation \eqref{eqn:AdjointGradient}, we adopted the Quasi-Newton iterative algorithm to reconstruct the forcing parameters.
The initial guess is a constant forcing $f=\alpha_0$, i.e., $\alpha_1= \alpha_2 =\ldots = \alpha_6=0$.
We start from initial guesses with $\alpha_0$ ranging from 0 to 4 and arrive at different solution of the forcing that drives the cost function sufficiently small, namely $J<10^{-6}$. 
\begin{figure}
    \centering
    \includegraphics[width=\textwidth]{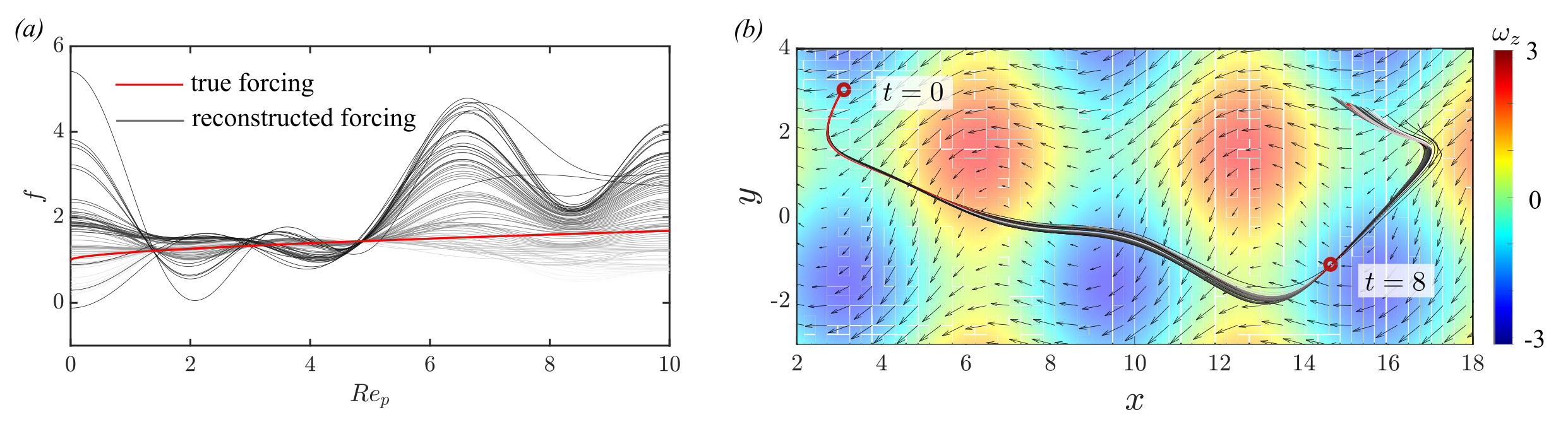}
    \caption{(a) Collection of reconstructed forcing starting from different initial guesses $f = \alpha_0 \in [0,4]$, darker color marks higher values of $\alpha_0$ in the initial guess. Red line marks the true forcing. (b) Top view of the collection of reconstructed particle traces. The locations of the particle at the initial $t=0$ and observation time $t=8$ are known and marked by the red circle. Colors of the particle traces correspond to the ones in (a). Colored contours show the vorticity $\omega_z$ of the background flow field at $z = 0$.}
    \label{fig:reconstructedforcing}
\end{figure}
Results of various reconstructed forcing functions are plotted in figure \ref{fig:reconstructedforcing}$(a)$ in grey colors, darker color represents larger $\alpha_0$ in the initial guess and red line marks the true forcing function.
The collection of reconstructed forcing demonstrates the ill-posedness of the problem, where infinite number of possible solutions exist.
However, the particle trajectories from the reconstructed forcing are plotted in figure \ref{fig:reconstructedforcing}$(b)$ follow only slightly different paths that coincide at $t=8$ before they further deviate.
This also implies that even with large range of function $f$, the form of the forcing can only permit a limited region of arrival for the particle. 

\begin{figure}
    \centering
    \includegraphics[width=\textwidth]{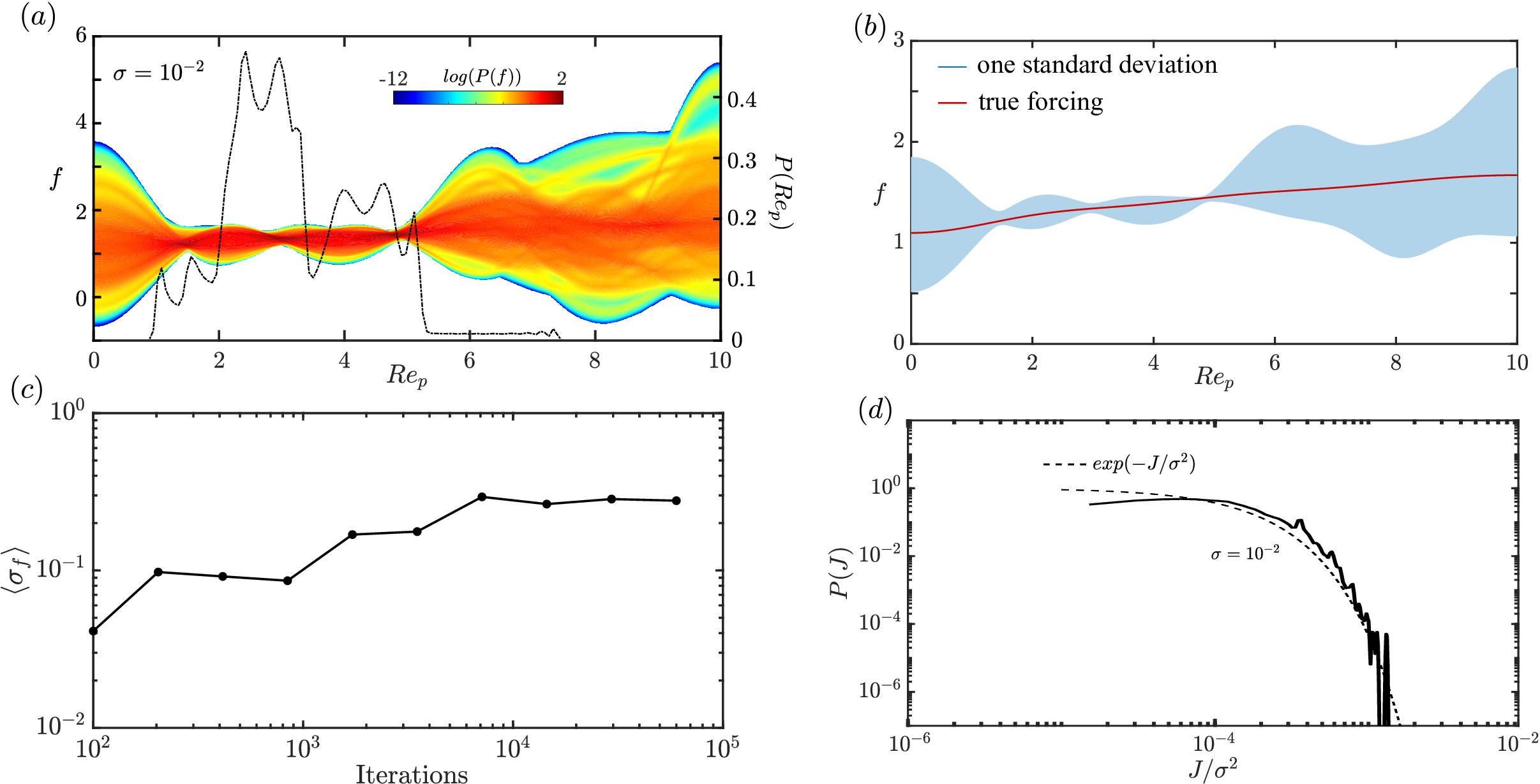}
    \caption{(a): Logarithmic of the Probability Density Function (PDF) of the forcing $f$ during HMC sampling. 
    % From top to bottom different level of noise tolerance, $\sigma = \{0.01,0.001,0.0001\}$ are used.
    The true forcing function is marked by the black line.
    (b) Statistics of the PDF from forcing estimation are plotted. Dashed lines represent the mean (expected values) of the forcing, while the shaded region marks forcing function within one standard deviation. The true forcing is shown by the red solid line. 
    %Darker colors represent less uncertainty in the observation.
    (c) Growth of the averaged standard deviation of the forcing distribution as more iterations are performed. 
    (d) Probability distribution of the cost function during HMC after the burn-in period. For different level of observation uncertainty $\sigma$, the theoretical results are also shown in dashed lines.}
    \label{fig:UQ}
\end{figure}
To further quantify the uncertainty of the forcing function $f$, the HMC algorithm is adopted.
The noise level $\sigma$ is the key to the efficiency of the algorithm. 
A noise level that is too large would result in unconstrained forcing function with unreasonable range.
On the other hand, a noise level that is too small would casue large curvature in the landscape of the cost function and cause difficulty for the sampling algorithm.
With the noise level $\sigma = 0.01$, the posterior distribution for the coefficients $\alpha_i$ of these modes is evaluated using HMC and shown in figure \ref{fig:UQ}.
%We focus on the observation of particle location at $t_m = 8$, with %different levels of noise in the observation, $\sigma = \{10^{-2},10^{-3},10^{-4}\}$.
%The initial guess of the forcing is a constant function, $f = 1$, corresponding to $\alpha_1 = 1, \alpha_i  = 0,i=2 \ldots 11$.

In order to obtain a reasonable starting point for the HMC algorithm, we start from the true forcing parameters. In real practice when the true forcing is unknown, this could be replaced by performing a few L-BFGS iterations to update the forcing coefficients from the initial guess until we reach the desired level of tolerance. 
This procedure, in lieu of the traditional burn-in period, is more efficient in term to get near the true solution and serves as a fair starting point for HMC algorithm.

The posterior probability distribution of the forcing is plotted as colored contours in figure \ref{fig:UQ}$(a)$. 
%The results, from the top panel with $\sigma = 0.01$ to the bottom panel with $\sigma = 0.0001$ clearly shows the trend of less uncertainty in the determined forcing when the uncertainty in the observation decreases.
The mean and standard deviation of the reconstructed forcing are plotted in figure \ref{fig:UQ}$(b)$.
The forcing reconstruction is much more accurate in the range $1 <Re_p< 5$ than at other particle Reynolds numbers. 
Compared with the history of $Re_p$ plotted in the background of figure \ref{fig:UQ}$(a)$, it is apparent that the $Re_p$ with low uncertainty of the forcing coincide with the high probability of occurrence along the trajectories.

The distribution of the cost function during the HMC sampling is plotted in figure 
\ref{fig:UQ}$(d)$, where the dashed lines marks the theoretical results of $P(J) \sim \exp (-J/\sigma^2)$.
The mismatch between the HMC and the theoretical prediction is due to the difficulty of sampling near the optimal parameters for high-dimensional problems due to low volume fraction interior of the typical set \citep{betancourt2017conceptual}.

\subsection{Homogeneous Isotropic Turbulence}
The algorithm is also tested in three-dimensional homogeneous isotropic turbulence in a cubic domain $\Omega=[0, 2\pi]\times[0, 2\pi]\times[0, 2\pi]$. 
\begin{figure}
    \centering
    \includegraphics[width=\textwidth]{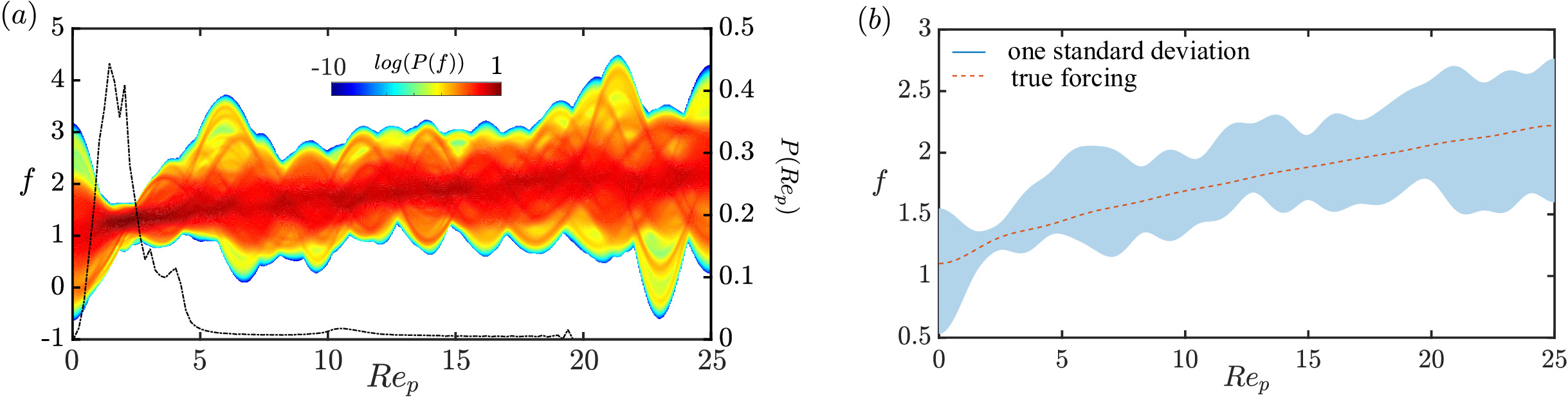}
    \caption{(a): Logarithmic of the Probability Density Function (PDF) of the forcing $f$ reconstructed using HMC. 
    % From top to bottom different level of noise tolerance, $\sigma = \{0.01,0.001,0.0001\}$ are used.
    The true forcing function is marked by the black line.
    (b) Statistics of the PDF from forcing estimation are plotted. Dashed lines represent the mean (expected values) of the forcing, while the shaded region marks forcing function within one standard deviation. The true forcing is shown by the red solid line. Darker colors represent less uncertainty in the observation.
    (c) Probability distribution of the cost function during HMC after the burn-in period. For different level of observation uncertainty $\sigma$, the theoretical results are also shown in dashed lines.}
    \label{fig:UQ_HIT}
\end{figure}
The flow fields are obtained with an in-house discontinuous Galerkin compressible DNS solver (\cite{klose2020assessing} and references therein).
For the definition of the initial condition of the flow field we adopted the one used in~\cite{jacobs2005validation,dominguez2022closed}.
The particle is initialized at location $(x_p,\ y_p,\ z_p)^\top=(\pi,\ \pi,\ 2\pi/3)^\top$ with velocity $(u_p,\ v_p,\ w_p)^\top=(0,\ 0,\ 1.5)^\top$.
The Reference Reynolds number is $Re_\infty=2,357$, the Stokes number is unity $St=1$, the particle density ratio is $\rho_p=250$ and the non-dimensional particle diameter $d_p=0.0055$. The observation time here is selected to be $t = 6$, where the particle travel for six Stokes time units being the time scales of changes in the flow on the same order than those on the particle phase ($St=1$).

The results of uncertainty quantification using 15 Fourier modes are shown in figure \ref{fig:UQ_HIT}. The uncertainty of the forcing function $f$ is the smallest for the $Re_p$ with the most occurrence along the particle trajectory.
This result is very similar to that in the ABC flow.
The underlying reason is that as the particle move faster than the ambient fluid, the momentum effect becomes more and more important. For large $Re_p$, inertia plays a vital role and the effect of forcing become nearly negligible, rendering the determination for large $Re_p$ much more challenging.
The probability of occurrence for different Reynolds number, $P(Re_p)$ along the trajectories during the HMC sampling are also plotted in figure \ref{fig:UQ}$(a)$.

%%%%%%%%%%%%%%%%%%%%%%%%%%%%%%%%%%%%%%%%%%%%%%%%%%%%%%%%%%%%%%%%%%%%%%%%%%%%%%%%
%							Conclusion
%%%%%%%%%%%%%%%%%%%%%%%%%%%%%%%%%%%%%%%%%%%%%%%%%%%%%%%%%%%%%%%%%%%%%%%%%%%%%%%%
\section{Conclusion}
A data assimilation framework is established, to estimate the forcing of particles as a function of the relative velocity $a=|\boldsymbol{u}_p - \boldsymbol{u}|$ based on sparse, noisy measurements of the arrival location of the particle.
The optimization framework relies on the adjoint dynamics of the particles, which yields the same Lyapunov exponent as the forward dynamics. 
Furthermore, a Hamiltonian Markov Chain is adopted for uncertainty quantification of the forcing, when measurements are subject to Gaussian noise.

The algorithm is demonstrated in ABC flow and isotropic turbulence with the Stokes number of unity.
The algorithm can efficiently identifies forcing functions to accurately guide particles to the observed arrival locations. 
The uncertainty quantification from HMC demonstrates the ill-posedness of the problem when observations has noise.
The accuracy of the forcing dependents on the history of particle Reynolds number along the trajectory. 
In both flows, the forcing for $1<Re_p<5$ can be more accurately determined than other Reynolds numbers. 

The current study is the first step for data assimilation in particle forcing determination using the adjoint particle dynamics and could lead to better understanding of kinematics for particle-laden flows from experimental data, and advance the data-based development for reduced-order model for particle-flow interactions. 
Possible future research includes extending the framework to multiple particles with and without labels, and incorporating the hydrodynamic memory.

%\bibliographystyle{jfm}
%\bibliography{jfm}
%Use of the above commands will create a bibliography using the .bib file. Shown below is a bibliography built from individual items.

\bibliographystyle{jfm}
\bibliography{Manuscript}

\end{document}